\documentclass[preprint,twocolumn,tightenlines,10pt,prl]{revtex4-1}%
\usepackage{amsfonts}
\usepackage{amsmath}
\usepackage{amssymb}
\usepackage{graphicx}%
\usepackage{color}
\usepackage{soul}
\usepackage{ulem}

\providecommand{\U}[1]{\protect\rule{.1in}{.1in}}
\newcommand{\quotes}[1]{``#1''}
\newcommand{\paren}[1]{\left ( #1 \right )}

\newcommand{\B}[1]{\textcolor{blue} { #1}}

\newcommand{\myequationn}[1]{\begin{align*} #1 \end{align*}}

\begin{document}

\title{Breaking Chiral Symmetry with Microfluidics}
\author{L. L. A. Adams$^*$ and S. A. Ocko${\dagger}$}
\affiliation{$^*$School of Engineering and Applied Sciences, Harvard University, Cambridge, MA 02138 and $^\dagger$Applied Physics, Stanford University, Stanford, CA 94305.}


\begin{abstract}
A robust route for the biased production of single handed chiral structures
has been found in generating non-spherical, multi-component double emulsions using microfluidics. The specific type of handedness is determined by the final packing geometry of
four different inner drops inside an ultra-thin sheath of oil. Before three dimensional
chiral structures are formed, the quasi-one dimensional chain re-arranges in two
dimensions into either checkerboard or stripe patterns. We derive an
analytical model predicting which pattern is more likely and assembles in
the least amount of time. Moreover, our \textit{dimensionless} model accurately predicts
our experimental results and is based on local bending dynamics, rather than global surface energy
minimization. {This better reflects the underlying self-assembly process which will not,  in general, reach a global energy minimum}. In summary, using glass microfluidic techniques for channeling aqueous fluids through narrow orifices of multi-bore injection capillaries while
encapsulating these fluids as drops inside an ultra-thin sheath of oil is
sufficient to produce single-handed chiral structures. 
\end{abstract}
\volumeyear{year}
\volumenumber{number}
\issuenumber{number}
\received[Received text]{date}

\revised[Revised text]{}

\startpage{1}
\endpage{102}
\maketitle

To understand the importance of single handed chirality, one has to look no
further than biomolecules that give us life \cite{Blackmond} - \cite{Jafarpour}. The amino acids and sugar
molecules that are building blocks of proteins and DNA are all homochiral 
and one of the great mysteries in understanding the origin of life is how
these enantiomers of an exclusive handedness came into existence \cite{Cronin} - \cite{Breslow}. The quandary
arises because chemical reactions produce racemic structures, that is, equal
amounts of left and right handed chiral molecules \textit{unless} painstaking
care is taken by introducing a chiral catalyst into a reaction flask.

In this Letter, we report a strategy for breaking left - right symmetry by
biasing the production of sub-millimeter-size chiral structures \cite{Chaikin} towards one type of
handedness.  Using glass microfluidic devices \cite{Utada}, the chiral structures we generate are
multi-component double emulsions, drops nested inside of drops. For generating chiral macro-``molecules" with this strategy, two conditions need to be met. One
condition is that the intrinsic spherical shape of the double emulsion
structure needs to be altered into something molecular-like, such as the shape
of configurational isomers \cite{manoharan}, \cite{Garstecki}. The second condition is encapsulation of at
least four different inner drops \cite{Zhao}, \cite{Lee} of either different sizes or different compositions.\ %

%TCIMACRO{\FRAME{ftbpFU}{3.4497in}{1.9623in}{0pt}{\Qcb{\QTR{bf}{Production of
%multi-component dimers and trimers.} a.) Schematic of microfluidic device for
%generating non-spherical double emulsions. Flow directions are indicated with
%dark arrows.The middle fluid is represented by the dark line on the outside of
%the injection capillary. b.) and c.) Optical microscope images of dimer and
%trimer generation in a microfluidic device using a dual bore injection
%capillary. The inner water drops, dyed red and blue, appear as white and grey
%colored drops, respectively, in the image. They are encapsulated in a thin
%sheath of oil that is barely discernable. For b.) Q$_{red}$ = 100 $\mu l/hr$
%and for c.) Q$_{red}$ \ = 500 $\mu l/hr.$ The flow rates of the blue inner
%fluid, Q$_{blue}$ = 100 $\mu l/hr$ , the middle fluid, Q$_{oil}$ = 300 $\mu
%l/hr,$ and outer fluid, Q$_{aq}$ =30,000 $\mu l/hr$ are constant in both
%cases. The scale bars are 100 $\mu m.$}}{}{fig1_best.pdf}%
%{\special{ language "Scientific Word";  type "GRAPHIC";
%maintain-aspect-ratio TRUE;  display "USEDEF";  valid_file "F";
%width 3.4497in;  height 1.9623in;  depth 0pt;  original-width 3.4039in;
%original-height 1.9242in;  cropleft "0";  croptop "1";  cropright "1";
%cropbottom "0";
%filename '../figuresnonspherical/fig1_best.pdf';file-properties "XNPEU";}}}%
%BeginExpansion
\begin{figure}
[ptb]
\begin{center}
\includegraphics[
width=\columnwidth
]%
{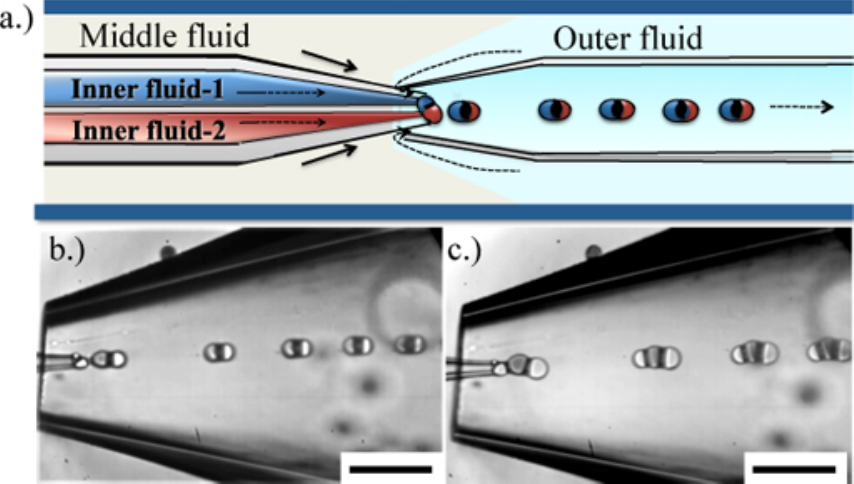}%
\caption{(Color online) \textbf{Production of multi-component dimers and trimers.} a.)
Schematic of microfluidic device for generating non-spherical double
emulsions. The middle fluid is
represented by very thin dark line on outside of injection capillary. b,
c.) Optical microscope images of dimer and trimer generation in a
microfluidic device using a dual bore injection capillary. Inner water
drops, dyed red and blue, appear as white and grey colored drops,
respectively. They are encapsulated in a barely discernable thin sheath of oil. For b.) Q$_{red}$ = 100 $\mu l/hr$ and for c.)
Q$_{red}$ \ = 500 $\mu l/hr.$  Flow rates of blue inner fluid,
Q$_{blue}$ = 100 $\mu l/hr$ , middle fluid, Q$_{oil}$ = 300 $\mu l/hr,$
and outer fluid, Q$_{aq}$ = 30,000 $\mu l/hr$ are constant in both cases. Scale bars are 100 $\mu m.$}%
\end{center}
\end{figure}
%EndExpansion
%

%TCIMACRO{\TeXButton{B}{\begin{figure*}}}%
%BeginExpansion
\begin{figure*}%
{\begin{center}
%\includegraphics[
%height=4.3794in,
%width=5.9776in
%]%
\includegraphics[
width = \textwidth]
{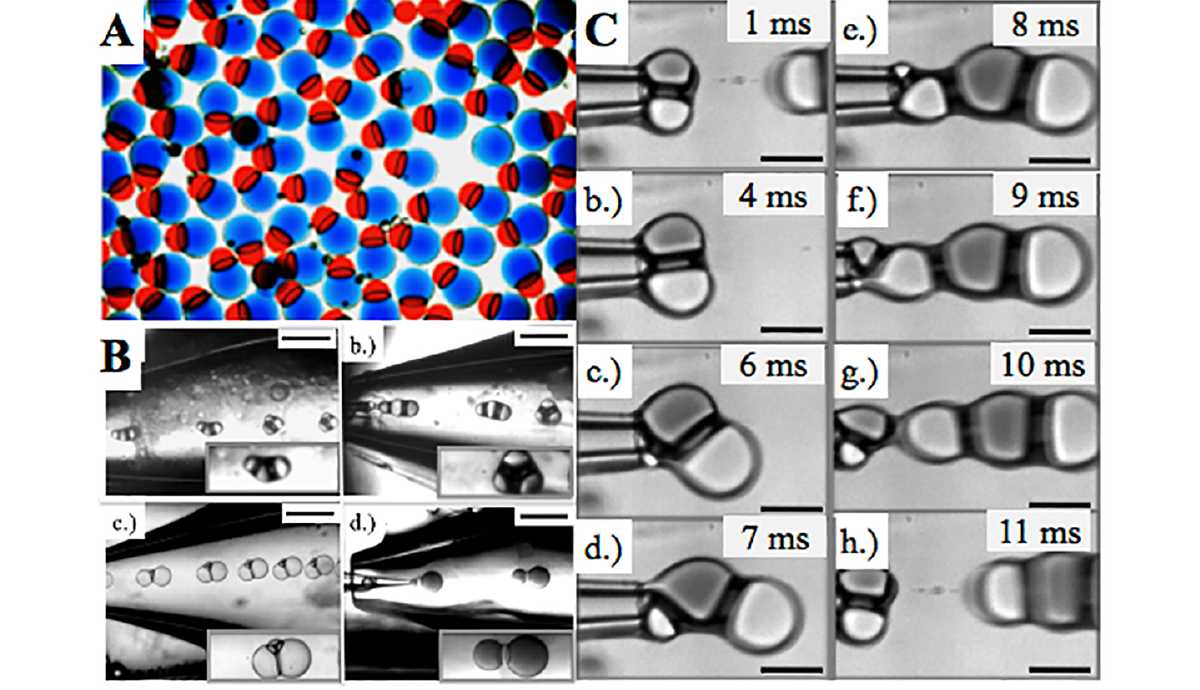}%
\\
FIG.\:2. (Color online) \textbf{\;Control and sequence of two different inner droplets
inside a limiting volume of oil.} A. Optical microscope image of dimers with
one red and one blue inner drop. B. Four different trimers. Trimers contain three different inner
drops inside an indiscernible ultrathin oil layer. $Insets:$ Magnified images
of individual double emulsions. Scale bars are 200 $\mu$m. C. Time Sequence: Trimer
Generation From a Dual Bore Capillary. a.) and b.) First two
drops are connected at the nozzle to their respective orifices by an
ultra-thin layer of oil. c.) Light colored drop pinches off first from its
orifice since its flow rate is 5x that of dark drops. d.) Another
light colored drop emerges from its orifice. e.) First two drops begin to
align with outer flow as they move away from nozzle while still
attached to a third drop. A fourth drop appears. f.) Third inner drop
begins to pinch off from nozzle while fourth drop becomes larger. g.)
Trimer begins to pinch-off. Last drop is smaller than
previous two, having spent less time at nozzle because a greater drag force pulls on it from
first two drops and continuous phase. h.) Trimer pinches
off and is free to move in the flow. Scale bars denote 30 $\mu$m.
\end{center}}%
%EndExpansion%
%TCIMACRO{\TeXButton{E}{\end{figure*}}}%
%BeginExpansion
\end{figure*}%
%EndExpansion
%

%TCIMACRO{\FRAME{ftbpFU}{3.4497in}{2.3964in}{0pt}{\Qcb{a.) Schematic of
%$\theta,z$ \ used in our analytical model. The colored drops are
%representative of the center of each drop. b.) The eigenvectors and
%eigenvalues for checkerboard and stripe patterns.}}{}{schematic2.pdf}%
%{\special{ language "Scientific Word";  type "GRAPHIC";
%maintain-aspect-ratio TRUE;  display "USEDEF";  valid_file "F";
%width 3.4497in;  height 2.3964in;  depth 0pt;  original-width 3.4039in;
%original-height 2.3557in;  cropleft "0";  croptop "1";  cropright "1";
%cropbottom "0";
%filename '../figuresnonspherical/schematic2.pdf';file-properties "XNPEU";}}
%}%
%BeginExpansion
\setcounter{figure}{2}
\begin{figure}
[ptb]
\begin{center}
\includegraphics[
width=\columnwidth
]%
{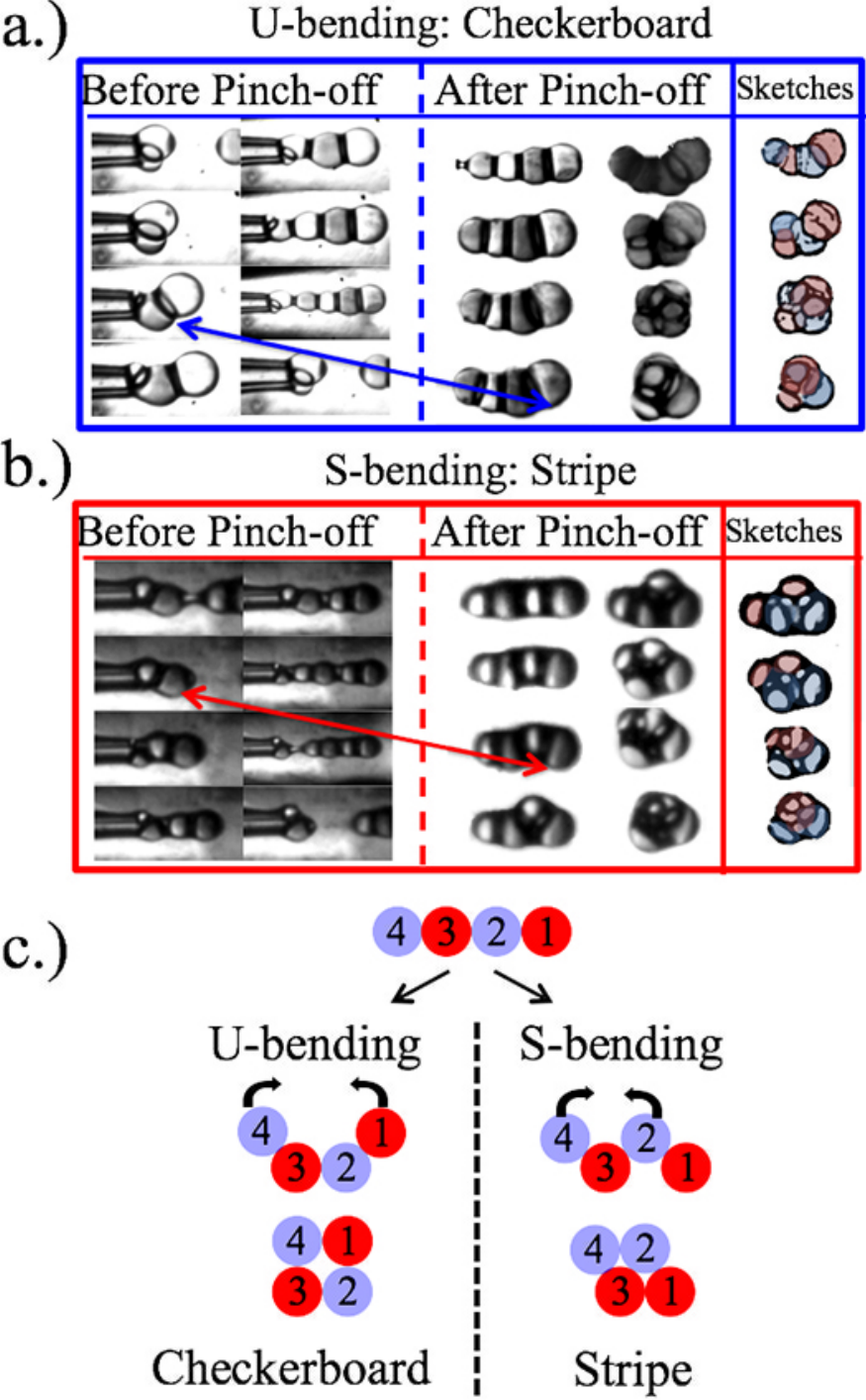}%
\caption{(Color online) Two different self assembly routes for generating chiral double emulsions both before and after pinch-off from a dual bore injection capillary. Last column in each set of images contains sketches of the final stages of the self-assembly process.  While re-arranging in 2D, drops form either  a.) Checkerboard pattern or  b.) Stripe pattern. c.) Illustration of U-bending and S-bending. Note how  U-bending leads to a checkerboard pattern while S-bending will lead to a stripe pattern.}%
\end{center}
\end{figure}

\begin{figure}
[ptb]
\begin{center}
\includegraphics[width=\columnwidth]
{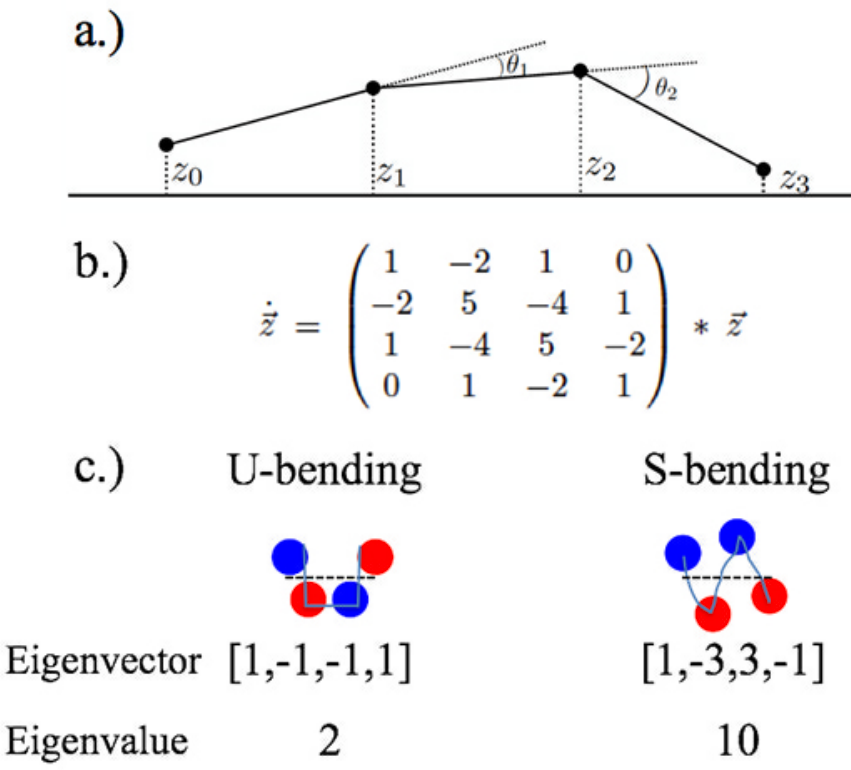}%
\caption{(Color online)a.) Schematic of $\theta,z$ \ used in our analytical model, b.) the Jacobian matrix , and c.)
eigenvectors and eigenvalues representing checkerboard and stripe patterns.}%
\end{center}
\end{figure}
%EndExpansion
We form non-spherical double emulsions by encapsulating multiple drops inside
an ultrathin shell using a single step emulsification technique. This
technique requires a custom designed multi-bore injection glass capillary in our
microfluidic device \cite{Adams}. To operate such a device, for example, a dual bore
injection capillary device, we pump at least four different fluids through the
device as seen in the schematic of Fig. 1 a. For double emulsions with just
two different inner drops, we use aqueous fluids containing red and blue dyes
as encapsulants. {These inner drops are stabilized by both surfactants and an ultrathin shell of the middle phase}

By fixing flow rates of middle
$Q$$_{oil}$,  continuous $Q$$_{aq}$, and one of the inner phases, two
different droplet configurations emerge. \cite{flow} For example, at high flow rates of
continuous phase, and low flow rate of
middle phase{,} either all dimers or all trimers
are produced containing two distinct components. Difference between dimer
and trimer production is increased flow rate of one of the inner fluids while keeping the other inner phase fixed
as seen in Fig. 1 b and 1 c.

To demonstrate the control we have, we generate a large number of
monodispersed double emulsions shaped as dimers containing two different sizes
of inner drops as shown in the optical microscope image in Fig. 2 A. To
illustrate the range of non-spherical geometries \cite{Cho} - \cite{Tabeling} possible with this technique,
we show four variations of trimers in Fig. 2 B. The four trimer variations are
each generated with a different device indicating the overall robustness of
our technique. The ultrathin sheath of oil is not observable in these figures,
as it is approximately 1000 nm or less \cite{thinshell}. Moreover, these
interfaces are highly stable due to thinness of the shell and addition
of surfactants in the oil preventing inner drops from merging with each
other or the continuous fluid.

{The mechanism for self-assembling non-spherical double emulsions, whether dimers, trimers or tetrahedral structures, is purely
deterministic. \cite{random}  For example, if the process was stochastic we would expect that
a red-blue-red trimer would eventually be followed by a blue-red-blue trimer;
instead, red-blue-red trimers are always followed by more red-blue-red trimers,
thus skipping a blue drop. This has the technical advantage of achieving monodispersity since
 the first two inner drops, red (light) and blue (dark) straddle the nozzle as seen in parts a) and b) of
Fig. 2 C. While these two coupled drops pivot, a third drop
emerges as shown in parts c) - e) in Fig. 2 C. \  When the shearing force from the continuous
phase overcomes the interfacial tension force \cite{Studart}, \cite{Erb}, the trimer begins to partially detach
%to break off containing a third drop smaller than
%the first two inner drops 
as seen in Fig. 2 C\thinspace\ f) and g)  until it completely detaches as seen in Fig. 2 C\thinspace\ h). }
%\Sam{De-duplication looks good. }

%As the trimer aligns with the flow,
%there is a slight curvature which is set before pinch-off and establishes the
%dynamics after pinch-off as the trimer minimizes its surface energy.

{
We now focus on dynamics of four inner drops, the smallest number of different drops that can form chiral
structures. As in the case of trimers, encapsulated drops
alternate between red and blue when exiting the capillary  to form a quasi-one dimensional chain.
With images taken from a high speed camera and optical microscope, we track
the motion of droplets before and after pinching-off from orifices of a
dual bore capillary as seen in Fig. 3 a and b. \ Four different sizes of
inner drops emerge with the last drop in the sequence again the smallest.
After detaching from the orifice, encapsulated drops initially stay in a  quasi-one
dimensional configuration because of their high volume fraction and surrounding
flow of continuous phase. Then they bend into two-dimensional and three-dimensional configurations that reduce surface energy as shown in the `after pinch-off' images of Fig. 3 and b. Perturbations to the initial linear configuration lead to one of two possible
instabilities:  drops can bend in a \quotes{U} shape forming a checkerboard pattern or an \quotes{S} shape to form a stripe pattern as illustrated in Fig. 3 c and shown {in the data} of Fig. 3 a and b. }\

{We develop an analytical
model to predict which two-dimensional arrangement, checkerboard or stripe,
is most prevalent and which bending mode requires the least amount of time. In our \textit{dimensionless} toy model,  each  drop, identical in size but alternating in color, is parametrized with a height $z_{i}$, and interacts with its nearest and next-nearest neighbors through a capillary term that depends only on bending angles as illustrated in Fig. 4 a.  Each bending angle, two angles for four droplets, is defined as the angle circumscribed by the centers of three neighboring drops.  In the limit of small angles, the bending angle is expressed as:\  $\theta_{i}=\frac{2z_{i}-z_{i+1}-z_{i-1}}{\textit{radius}}$  with the radius of drops taken to be 1. Since we are interested in the small-angle limit, we approximate the \textit{dimensionless} bending energy to be quadratic: \ $  E= -\frac{1}{2} \paren{ \paren{\theta_{1}}^{2} + \paren{\theta_{2}}^{2}}$. \cite{energy}
%
%
%In our toy model, we parameterize four
%equally sized drops by identifying and associating each drop's center with a
%height, z$_{i}$ as illustrated in Fig. 4 a.
}\

{Since the system is at low Reynolds number, $Re$ $\approx$ 0.5 \cite{Reynolds}, droplet re-arrangement rates will be linear with the capillary force. Assuming the system takes a steepest descent 
\cite{descent}, the bending rate can be expressed as the potential energy gradient $\dot{z}_{i} = -\frac{dE}{dz_{i}}$ \cite{hessian}. From this, we can construct a Jacobian matrix \textit{M} where $\dot{\vec{z}} = -\frac{dE}%
{d\overrightarrow{z}}=M\,\overrightarrow{z}$ as shown in Fig. 4 b. Solving this gives two non-trivial eigenvectors representing \quotes{U} and \quotes{S} - bending:  (1,-1,-1,1) and (1,-3, 3,-1) for eigenvalues 2 and 10, respectively, as shown in Fig. 4  c).}\

%
%{
%Since the system is at low Reynold's number, Re
%%TCIMACRO{\TEXTsymbol{<} }%
%%BeginExpansion
%$<$
%%EndExpansion
%1, the capillary force, which drives the drop's re-arrangements, is
%proportional to velocity or the re-arrangement rate. This rate can be
%expressed as the potential energy gradient, $\frac{dE}{dz_{i}}$.\ From this
%expression, we can construct a Hessian matrix, M, where $\frac{dE}%
%{d\overrightarrow{z}}=M\,\overrightarrow{z}.$ Assuming the system takes a
%steepest descent, it does not exactly because of viscous forces, the
%re-arrangement rate, $\frac{d\,\overrightarrow{z}}{dt}\approx
%-M\,\overrightarrow{z}$. Solving this for the eigenvalues and eigenvectors of
%the Hessian matrix -$M$ \ gives two non-zero eigenvalues, 2 and 10 which have
%eigenvectors (1,-1,-1,1), U-bending, and (1,-3,3,-1), S-bending, respectively
%as shown in Fig. 4 b.
%}

{Our model predicts that configurations with \quotes{S} - bending (stripe)
will fold 5 times faster than \quotes{U} - bending (checkerboard). Using image analysis techniques, we calculate the temporal evolution of  overall length/initial length for three double emulsions with a stripe pattern and three double emulsions with a checkerboard pattern as shown in Fig. 5 a. It should be noted that all subsequent double emulsions re-arrange exactly the same way. The initial length is the length of the linear configuration immediately after detaching from the orifice.} \

{Remarkably \cite{remark}, we find good agreement between observed bending time ratios and our model: in experiments, \quotes{S} - bending double emulsions are faster than  \quotes{U} - bending by $\sim$ 5.6 times.  Difference between theory and experiment is likely due to a combination of fluid mechanical considerations, where due to lubrication effects and low
Reynolds number interactions, the system will take
something other than a steepest descent, and non-linear terms will come into play when displacements become larger. \cite{remark2}}\
%BeginExpansion
\begin{figure}
[ptb]
\begin{center}
\includegraphics[
width= \columnwidth
]%
{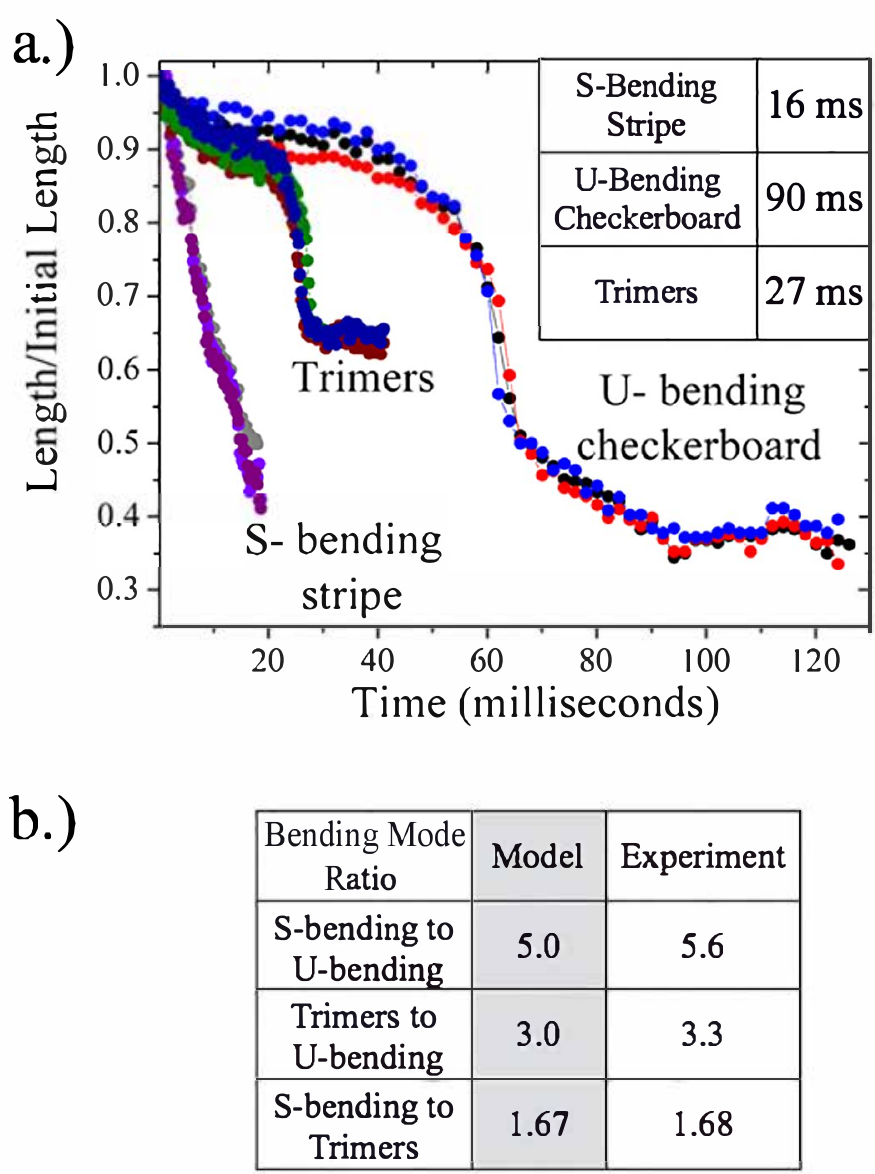}%
\caption{ (Color online) a.) Overall length length/initial length for 3 checkerboard, 3 stripe, and 3 trimer double emulsions as a function of time. Double emulsions which generate stripe patterns assemble 5.6 times faster than double emulsions which generate checkerboard patterns. Insert:  A summary of re-arrangement times for each bending mode. b.) Table comparing the ratio of three different bending modes, the ratio of eigenvalues, to the ratio of re-arrangement times from experiments.}
\end{center}
\end{figure}
%EndExpansion
%
%
%
%{This model predicts that the stripe instability will be more prevalent than
%the checkerboard instability, and that configurations with stripe patterns
%will fold \symbol{126}5 times faster. This is similar to the observed ratio
%\ of 4.5 times between stripes over checkerboard folding times as shown in
%Figure 5. The difference between the model and experimental results is likely
%due to a combination of fluid mechanical considerations, where due to low
%Reynolds number interactions and lubrication effects, the system will take
%something other than a steepest descent and nonlinear terms that occur when
%displacements become larger.
%}%
%TCIMACRO{\TeXButton{B}{\begin{figure*}}}%
%BeginExpansion
\begin{figure*}%
%EndExpansion%
%TCIMACRO{\FRAME{itbpFU}{6.0623in}{2.7544in}{0in}{\Qcb{Figure 4. Two different
%self assembly routes for generating chiral double emulsions both before and
%after pinch-off from a dual bore injection capillary. The last column in each
%set of images contains sketches of the final stages of the self assembly
%process.}}{}{fig3_best.pdf}{\special{ language "Scientific Word";
%type "GRAPHIC";  maintain-aspect-ratio TRUE;  display "USEDEF";
%valid_file "F";  width 6.0623in;  height 2.7544in;  depth 0in;
%original-width 6.0027in;  original-height 2.7121in;  cropleft "0";
%croptop "1";  cropright "1";  cropbottom "0";
%filename '../figuresnonspherical/fig3_best.pdf';file-properties "XNPEU";}} }%
%BeginExpansion

%EndExpansion%
%TCIMACRO{\TeXButton{E}{\end{figure*}}}%
%BeginExpansion
\end{figure*}%
%EndExpansion

%
%With images taken from a high speed camera and optical microscope, we track
%the motion of droplets before and after pinching-off from the orifices of a
%dual bore capillary as seen in Fig. 3 B, C, and D. \ Four different sizes of
%inner drops emerge with the last drop in the sequence again the smallest.
%After detaching from the orifice, the encapsulated drops initially form a one
%dimensional chain because of their high volume fraction and the surounding
%flow of the continuous phase. A perturbation causes the sequence of drops to
%re-arrange to minimize their surface energy.

{Our model also correctly predicts that \quotes{S}-bending will be more prevalent than \quotes{U}-bending as an initial S-perturbation will quickly outgrow an initial U-perturbation of the same size. However, even though experiments confirm this, our model predicts this without capturing initial conditions which also play a vital role in the final outcome of stripes or checkerboards. %Bending modes are experimentally deterministic, not probabilistic, since they are  device and flow rate dependent. 
Final configurations depend on initial angles %at which the first two drops pinch-off 
as indicated by the arrows in Fig. 3 a and b. No two devices are identical: a device that produces an initial state with a \textit{very} strong U-perturbation will overpower the dynamical bias towards S-bending, giving U-bent (checkerboard) final configurations. 
%because ofmodifications made to the glass injection and collection capillaries before
While the overwhelming majority of devices produce stripes, some devices will generate checkerboards. The key feature, though, is that once a configuration is generated, subsequent double emulsions produced will bend in \textit{exactly} the same way \cite{youtube}. This is also evident with dimers and trimers. }   

{To further validate our toy model, we compare re-arrangement rates of tetragonal structures, both stripes and checkerboards, to trimers as shown in Fig. 5 a. With trimers, the model gives only one eigenvalue, 6, corresponding to an eigenvector of (-1,2,-1)  \cite{trimer}. By evaluating the ratio of eigenvalues to the ratio of re-arrangement times, there is remarkable agreement between the model and experiment as seen in the values of table of Fig. 5 b.}

{As seen in videos, \cite{youtube}, all inner drops
re-arrange in exactly the same manner for a particular device and flow
rates. Videos show multi-component double emulsions which
break left-right symmetry by generating tetragonal double emulsions of only one type of handedness.
%The re-arrangement of inner drops is a consequence of how droplets emerge
%from the capillary injection tube and the restricted flow of oil sandwiched between 
%inner drops and their ultra-thin shell where lubrication dynamics plays a critical role. 
Moreover, in all experiments involving a dual bore injection capillary, we produce chiral structures of only one type of handedness. }

{One question worth addressing is why tetragonal structures do not form immediately at the nozzle. In our three dimensional device, hydrodynamic flow of the continuous phase plays a different and more complicated role than for two-dimensional PDMS devices where strings of drops are initially forced into linear configurations by the walls of the device before entering a wide chamber \cite{Garstecki}. In our work, increasing the shear force by increasing flow velocity of the continuous phase results in drops being stretched out into a nearly linear chain with the last drop anchored at the nozzle; this chain of drops is prevented from bending because of high fluid shear at the nozzle.  %  the flow velocity of the continuous phase results in the last drop anchored at the nozzle being prevented from bending because the high flow velocity stretches the drops out into a linear chain. 
 Once drops detach from the capillary's nozzle, the drops inside their thin elastic membrane are free to shuffle around as they travel to a region of lower fluid shear in the device.}\

{In summary, we have generated, using glass microfluidic techniques, linear chains of encapsulated drops which self assemble into chiral structures with the same handedness.  Our analytical model based on bending rates, rather than global energy minimization, presents a new approach for describing dynamics of self-assembling droplet structures. Understanding the dynamical behavior of these encapsulated inner drops through bending rates can allow the prediction and construction of more complicated, yet flexible, self-assembling structures than those presented here. Furthermore, if these structures can be made to respond to external triggers before changing their shape, a whole new range of applications in various industries, particularly the pharmaceutical and soft robotics industries, would be possible.}\

Acknowledgments: We gratefully acknowledge David Weitz's support for which without, this work would not be possible. {We also thank Piotr Garstecki, Ming Guo, Jan Guzowski, Tom Kodger, Stephan Koehler,  L. Mahadevan, Vinny Manoharan, Ian Morrison and Frans Spaepen for helpful suggestions and comments. This work was supported by the NSF (DMR-131026), the Harvard MRSEC (DMR-0820484),  Henry W. Kendall physics fellowship (S.O.), and the Karel Urbanek applied physics fellowship (S.O.).}\

\begingroup
\let\clearpage\relax

\endgroup
\clearpage


\begin{thebibliography}{99}                                                                                               %

\bibitem{Blackmond}D. G. Blackmond, Cold Spring Harbor Perspect. Biol., 2a002147 (2010).

\bibitem{Cronin}J. R. Cronin and S. Pizzarello, Science \textbf{275}, 951 (1997).

\bibitem{Rikken}G. L. Rikken and E. Raupach, Nature \textbf{405}, 932 (2000).

\bibitem{Lough}W. J. Lough and I. Wainer, \textit{Chirality in Natural and Applied Science}, Blackwell Publisher (2002).

\bibitem{Jafarpour}F. Jafarpur, T. Biancalani, and N. Goldenfeld, Phys. Rev. Lett. \textbf{115}, 158101 (2015).

\bibitem {Cronin}J. R. Cronin and J. Reisse, in:  M. Gargaud, ed. Lectures in Astrobiology, Vol. 1: Berlin: Springer-Verlag, 473-517 (2005).

\bibitem{Mason}S. F. Mason, Nature \textbf{311}, 19 (1984).

\bibitem{Breslow}R. Breslow, Tetrahedral Lett. \textbf{52}, 2028 (2011).

\bibitem {Chaikin} D. Zerrouki, J.  Baudry, D. Pine, P. Chaikin, and J. Bibette, Nature \textbf{455}, 380  (2008).

\bibitem {Utada}A. S. Utada, E. Lorenceau, D. R. Link, P. D. Kaplan, H. A.
Stone, D. A. Weitz, Science, \textbf{308}, 537-541 (2005).  

\bibitem {manoharan}V. N. Manoharan, M. T. Elsesser, and D. J. Pine, Science, \textbf{301}, 483 - 487 (2003).

\bibitem {Garstecki}J. Guzowski and P. Garstecki,
Phys. Rev. Lett., \textbf{14}, 188302 (2015).

\bibitem{Zhao}Y. Zhao, Z. Xie, H. Gu, L. Jin, X. Zhao, B. Wang, and Z. Gu, NPG Asia Materials \textbf{4}, e25 (2012).

\bibitem{Lee} S. S. Lee, A. Abbaspourrad, and S.-H. Kim, ACS Appl. Mater. Interfaces, \textbf{6} (2), 1294 (2014).

\bibitem {Adams}L. L. A.\ Adams, T. E. Kodger, S. -H. Kim, H. C. Shum, T.
Franke, and D. A. Weitz, Soft Matter, \textbf{ 41}, 10719-10724 (2012). 

\bibitem{flow} Heuristically we found: (a.) Increasing the flow of the continuous phase, fewer drops were encapsulated, (b.) Increasing the flow inner fluids, increased the number of encapsulated drops, and (c.) Increasing the flow of the middle phase alters the non-spherical shape to spherical drops inside of spherical drops.

\bibitem{Cho} Y.-S. Cho, G. -R. Yi, J. -M. Lim, S. -H. Kim, V. N. Manoharan, D. J. Pine, and S. -M. Yang, J. Am. Chem. Soc., \textbf{127}, 15968 (2005).

\bibitem {DLee}D. Lee and D. A. Weitz, Small, \textbf{5}, 1932-1935, (2009).

\bibitem {Thiele}J. Thiele and S. Seiffert, Lab on a Chip, \textbf{11}, 3188-3192, (2011).

\bibitem {Tabeling} B. Shen, J. Ricouvier, F. Malloggi, and P. Tabeling, Adv. Sci \textbf{3}, 1600012, (2016).

\bibitem {thinshell}S.-H. Kim, J. W. Kim, J.-C. Cho, and D. A. Weitz, Lab on a
Chip, \textbf{11}, 3162 (2011).

\bibitem {random} While the process \textit{after} the droplets pinch-off from the orifice is deterministic, the process of device fabrication is random. The known uncontrollable variables contributing to the uncertainty in the outcome of the dynamics are: the chemical surface treatment of the capillaries to make the surface hydrophobic or hydrophilic, wetting properties of the fluids to the capillaries, orifice diameters, the separation distance between the injection and exit capillaries, their alignment with respect to each other, and possible subtle angles of the orifices in how the capillaries are pulled from a commercial micro pipet controller.  We want to emphasize that once the chain of drops emerge from the orifice, no randomness is added to the system.


%\bibitem {Nisisako}T. Nisisako, S. Okushima and T. Torii, Soft Matter, 2005,
%\textbf{1}, 23-27.

%\bibitem {Romanowsky}M. B. Romanowsky, A. R. Abate, A. Rotem, C. Holtze, and
%D. A. Weitz, Lab on a Chip, 2012, \textbf{12}, 802-807.

%\bibitem {Geng}Y. Geng, P. Dalhaimer, S. Cai, R. Tsai, M. Tewari, T. Minko,
%and D. E. Discher, Nature Nanotechnology, 2007, \textbf{2}, 249-255.

%\bibitem {Champion}J. Champion, Y. Katare, and S. Mitragotri, J. of Controlled
%Release, 2007, \textbf{2}, 3-9.

%\bibitem {Panyam}J. Panyam, M. A. Dali, S.K. Sahoo, W. X. Ma, S. S.
%Chakravarthi, G. L. Amidon, R.J. Levy, V. Labhasewar, J. Controlled Release,
%2003, \textbf{92}, 173-187.

%\bibitem {Dunne}M. Dunne, O. I. Corrigan, Z. Ramtoola, Biomaterials, 2003, 21, 1659-1668.

%\bibitem {Tao}L. Tao, W. H, Y. Liu, G. Huang, B. Sumer, and J. Gao,
%Experimental Biology and Medicine, 2011, \textbf{236}, 20-29.

%\bibitem {Muro}S. Muro, C. Garnacho, J. A. Champion, J. Leferovich, C.
%Gajewski, E. Schuchman, S. Mitragotri and V. R. Muzykantov, Mol. Ther, 2008,
%\textbf{16}, 1450-1458.

%\bibitem {Donev}A. Donev, I. Cisse, D. Sachs, E. Variano, F. H. Stillinger, R.
%Connelly, S. Torquato, P. M. Chaikin, Science, 2004, \textbf{303}, 990-993.

%\bibitem {Kim}J. W. Kim, R. J. Larsen, D. A. Weitz, Adv. Mater.,
%2007,\textbf{19}, 2005 - 2009.

%\bibitem {Mitragotri}S. Mitragotri and J. Lahann, Nature Materials, 2009,
%\textbf{8}, 15-23.

%\bibitem {Decuzzi}P. Decuzzi and M. Ferrari, Biomaterials, 2006, \textbf{27}, 5307-5314.

%\bibitem {Lee}D. Lee and D. A. Weitz, Small, 2009, \textbf{5}, 1932-1935.

%\bibitem {Thiele}J. Thiele and S. Seiffert, Lab on a Chip, 2011, \textbf{11},3188-3192.

%\bibitem {Seiffert}S. Seiffert, Macromol. Rapid Commun, 2012, \textbf{33}, 1286-1293.

%\bibitem {Shin-Hyun}S.-H Kim, H. Hwang, C. H. Lim, J. W. Shim, and S.-M. Yang,
%Adv. Funct. Mater, 2011, \textbf{21}, 1608-1615.

%\bibitem {Choi}C.-H Choi, D. A. Weitz and C. -S. Lee, Adv. Materials, 2013,
%\textbf{25}, 2536-2541.

%\bibitem {Pannacci}N. Pannacci, H. Bruus, D. Bartolo, I. Etchart, T. Lockhart,
%Y. Hennequin, H. Willaime, and P. Tabeling, Phys. Rev. Lett., 2008,
%\textbf{101}, 164502-1 - 164502-4.

\bibitem {Studart}S. Barley, E. R Weeks, and K. Dalnoki-Veress, Eur. Phys. J. E \textbf{38}, 138 (2015).

\bibitem {Erb}R. M. Erb, D. Obrist, P. W. Chen, J. Studer, and A. R. Studart, Soft
Matter, \textbf{7}, 8757 (2011).


%\bibitem {shin-hyun2}S.-H. Kim, J. W. Shim, and S. -M. Yang, Ang.Chemie Int.
%Ed., \textbf{50,} 1171-1174, 2011.
\bibitem {energy}  The \textit{dimensional} bending energy can be expressed as: E = - $\gamma$   $\textit{l}^2$  ($\theta_1^2$ + $\theta_2^2$)
where $\gamma$ is the surface tension and \textit{l} is a characteristic length. For this expression to be valid, two important points must be satisfied: (a) \textit{l} must be less than the radius \textit{R} and (b) the distance \textit{z} must be less than the radius \textit{R}. For details, see Supplemental Material.


\bibitem {Reynolds} Re $= {\rho v L}/{\mu }$ where velocity for stripes is 0.006 m/s (Re $=$ 0.645) and velocity for checkerboards is 0.002 m/s (Re $=$ 0.337).

\bibitem {descent} Viscous forces will keep it from taking an \textit{exact} steepest descent.

\bibitem{hessian} The Jacobian matrix \textit{M} is the Hessian matrix.  For details, see Supplemental Material. 



\bibitem{remark}  Agreement between the toy model and experiment is remarkable since the assumptions in the toy model gloss over experimental details.  In the experiment, drops are not all the same size, the drops are never arranged in a truly one dimensional chain, there are more complex viscous interactions between droplets, etc. Even though these experimental features, which would be challenging to simulate, are not considered, the model does a strikingly good job at capturing the dynamics.Furthermore the non-monodispersity could be treated as a perturbation to the Jacobian matrix. However, since the eigenvalues for the non-perturbed Jacobian are so well separated with the S-bends 5 times faster than the U-bends, any perturbation to account for differences in sizes would not significantly alter the results. That is, to modify the eigenvalues, one would need a `super' perturbation. Thus our model also indirectly holds for polydispersed drops.

\bibitem{remark2} The distance \textit{z} in our model is not much less than the radius $R$.  Nonetheless, the model still captures the observed bending dynamics. 

\bibitem{youtube} Supplemental Movie. \B{ https://youtu.be/X1eCEKe-8no}

\bibitem{trimer}The trimer Jacobian is:
\myequationn{
M_{\text{Trimer}} =\;
   \begin{pmatrix} % or pmatrix or bmatrix or Bmatrix or ...
      1 &- 2 &1  \\
     - 2 & 4 &-2  \\
      1 & -2 &1 \\
   \end{pmatrix}
}


\end{thebibliography}
\end{document}

\title{Supplemental Material for \quotes{Breaking Chiral Symmetry with Microfluidics} }

\maketitle

\section{Fluids, Device Materials and Methods}

\subsection{Fluids}

For generating multi-compartment double emulsions, we use the following fluid
combinations as previously reported in \cite{Adams}: the outer phase is water
with 10 w/w \% poly (vinyl alcohol) (PVA; MW: 13000-23000 g/mol, 87-89\%
hydrolyzed, Sigma-Aldrich Co.), the middle phase contains kerosene with 8 w/w
\% PGPR-90 and 2 w/w/\% Abil-EM -90, and the inner phases contain water with
either Allura Red (Sigma-Aldrich), Toluidine
Blue (Spectrum), Wright Stain Blue (Sigma-Aldrich) dyes or without any dyes.
Before putting fluids that flow through the channels of the multi-bore capillary into syringes, the fluids are filtered
through 0.7 $\mu m$ filter paper.  This prevents
clogging of device channels.

\subsection{Preparation of devices}

Capillary microfluidic devices are fabricated with round dual, triple, and
quadruple bore injection capillaries and square capillaries purchased from
Atlantic International Technology (AIT glass). For example, the dual bore
glass capillary has an outer diameter of 1.55 -1.65 mm, bore sizes of 400
microns, and a septum of 0.635 mm.

The capillaries are tapered with a micropipette puller (Suter Instrument Co,
Model P-97) with additional tapering of the collection capillary achieved by
sanding the orifice with sand paper. Before anchoring the round capillaries
into the square capillary, their surfaces are dipped in one of two silanes for
a few minutes. To render the multi-bore injection capillary hydrophobic, we
dip the capillary into trimethoxy(octadecyl)silane (Sigma-Aldrich) ; to
render the collection capillary hydrophillic, we dip it into
2-[methloxy(polyethyleneoxyl-propyl] 9-12 trimethoxysilane (Gelest). The
injection capillary is either anchored slightly inside or outside the
collection capillary with epoxy.

\subsection{Device operation}

Each inlet of the device is connected to glass syringes using polyethylene
tubing (Scientific Commonities-PE5). The flow rates of fluids are controlled
using syringe pumps (Harvard PHD\ 2000 series). Typical flow rates used for
producing non-spherical drops are: outer flow rate, Q$_{aq}$,  which is between 10,000-30,000 $\mu l/hr$,
middle, Q$_{oil}$, and inner flow rates, Q$_{red}$ and Q$_{blue}$, which are between 100-500 $\mu l/hr$. The combined inner flow rates are about twice the middle phase's flow rate.

\subsection{Sample Characterization}

The production of non-spherical double emulsions is monitored in the
microfluidic device using an inverted optical microscope (DM-IRB, Leica)
fitted with a fast camera (Phantom 9, Vision Research). After fabrication, the
emulsions are imaged on a glass slide with an inverted microscope (Nikon
TE2000-E) equipped with a color digital camera (Nikon Sight DS-U1).

\subsection{Description of Chiral Double Emulsion Generation }

Before pinching off from the nozzle of the injection capillary, inner
drops of different sizes pack inside an ultrathin film of oil with alternating red and
blue colored aqueous drops. The four encapsulated aqueous drops, bound by a thin meniscus of oil,  proceed from the injection capillary in succession maintaining a quasi-linear
configuration of red-blue-red-blue inner drops.

\begin{figure}[htbp] %  figure placement: here, top, bottom, or page
   \centering
   \includegraphics[width=4in]{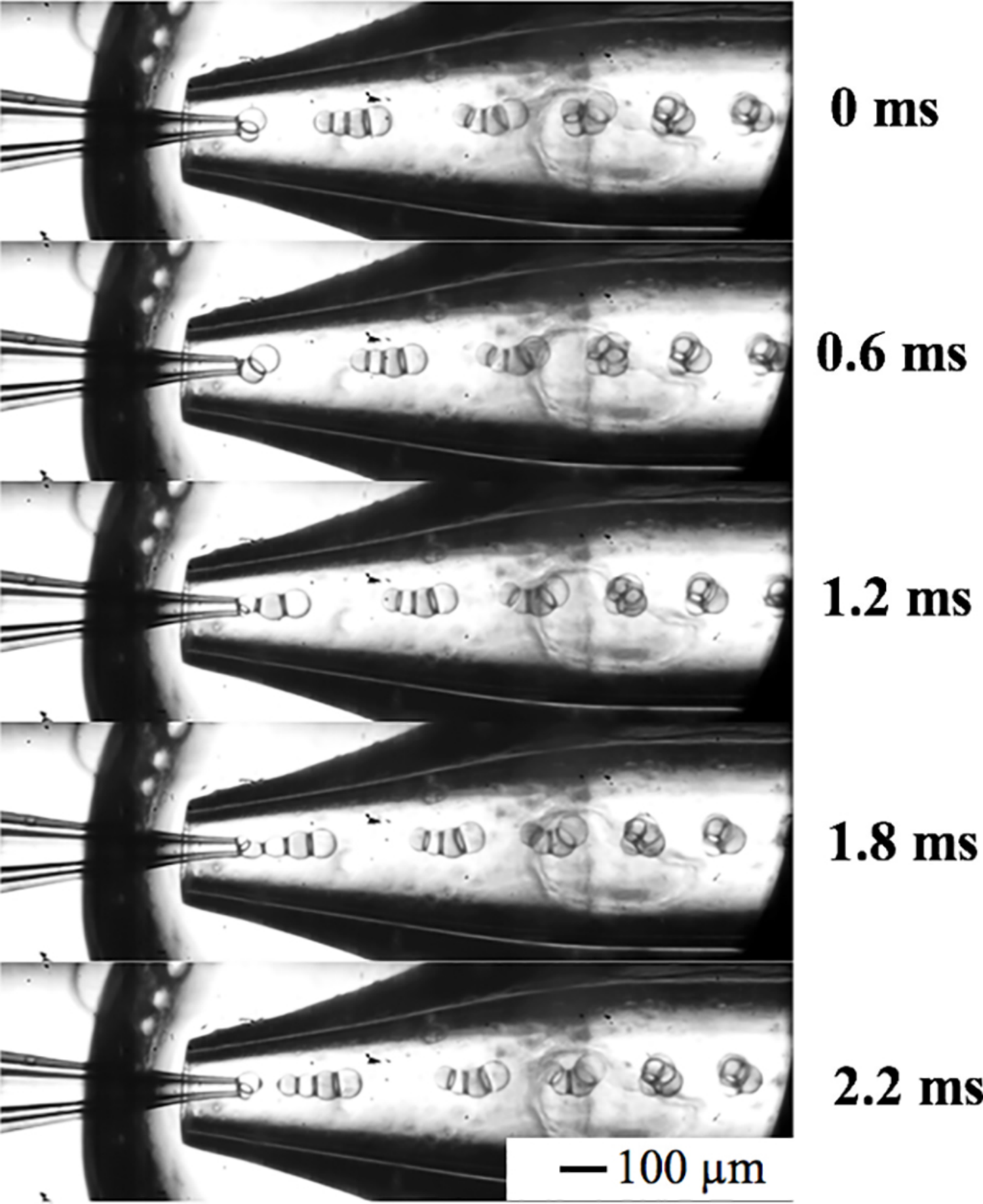} 
   \caption{Time sequence of U-bending. The scale bar is the same for all images.}
   \label{fig:example}
\end{figure}
After pinch-off, interfacial tension forces the linear configuration to fold
into compact structures. The evolution is recorded using a Phantom fast camera as seen in the time lapsed images of U-bending in Figure S1. Due to the system being at low Reynolds number, as the inner drops re-arrange, they do so in a completely deterministic manner. As the filament of encapsulated drops is never perfectly straight after pinch-off, how the drops exit the orifice of the injection capillary determines how they re-arrange downstream. Additionally, the inner drops do not coalesce with each other due to the thin
meniscus and surfactants between them. Moreover, each microfluidic device produces double emulsions which self assemble 
identically as shown in the time series of Figure S1.

\begin{figure}[htbp] %  figure placement: here, top, bottom, or page
   \centering
   \includegraphics[width=4in]{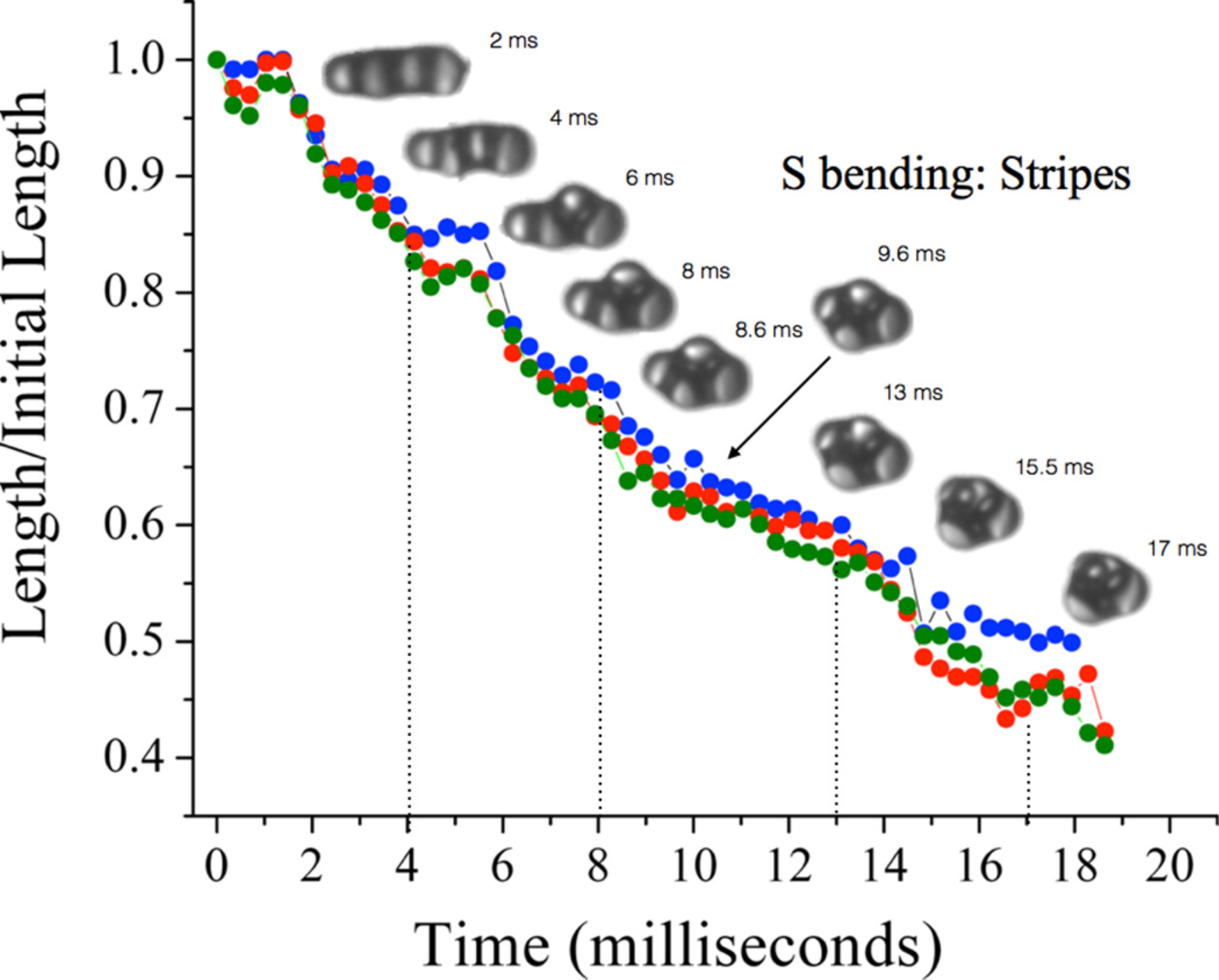} 
   \caption{S-bending with images of the double emulsions.}
   \label{fig:example}
\end{figure}

\begin{figure}[htbp] %  figure placement: here, top, bottom, or page
   \centering
   \includegraphics[width=4in]{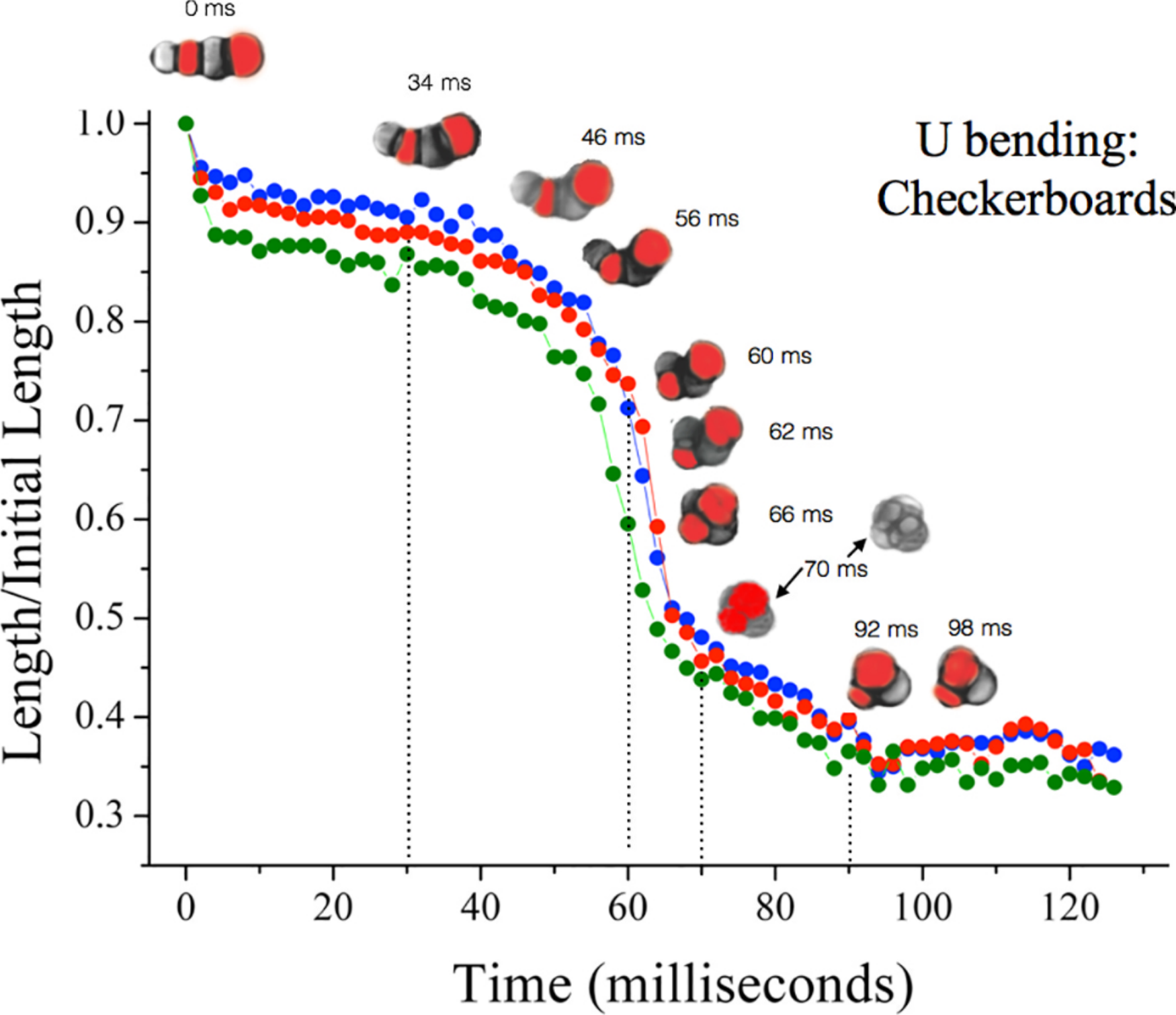} 
   \caption{U-bending with images of the double emulsions. Red drops are colorized to guide the eye.
%   Two drops are colored red to show explicitly the dynamics.
   }
   \label{fig:example}
\end{figure}

\begin{figure}[htbp] %  figure placement: here, top, bottom, or page
   \centering
   \includegraphics[width=5in]{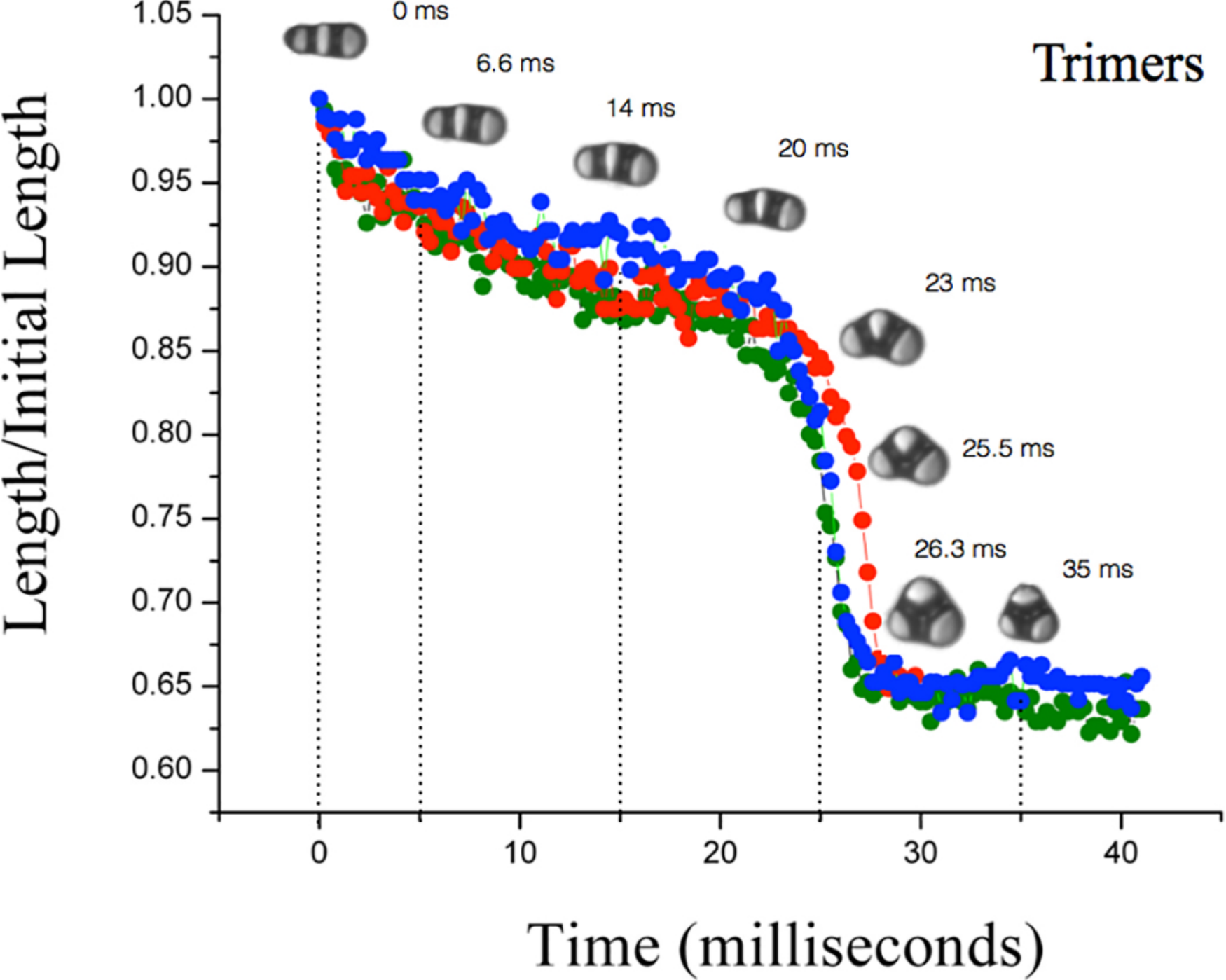} 
   \caption{Drops re-arranging into trimer configurations.}
   \label{fig:example}
\end{figure}

\subsection{Surface tension}
Interfacial tension between red water drops and kerosene with surfactants is 4.8 mN/m,
interfacial tension between blue water drops and kerosene with surfactants is 5.6 mN/m, and interfacial tension between kerosene and surfactants with the continuous aqueous phase is 3.8 mN/m.\\

%\subsection{Videos of devices with two bore injection capillaries}

%\B{Add a description of videos to be included with supplemental material}

%\subsection{Chiral structures using three and four bore injection capillaries}

%\B{Add a description of images to be included with supplemental material-use figure 6}
\clearpage
\newpage
\section{Derivation of Jacobian Matrices For Trimers and Quadramers}

%%%%%%%%%%%%%%%%

\subsection{Note on Hessians and Jacobians}
Consider a set of particles having positions $\vec{z}$. The energetic interaction force between particles is
 \myequationn{
% \textbf{F}_{\text{Interaction}}\paren{z}  = -\frac{d \text{E}_{\text{Interaction}}}{d \vec{z} }
 \interactforce \paren{z} = -\frac{d \interactenergy }{d \vec{z} }
%  \textbf{F}_{\text{Interaction}}\paren{z}  = -\frac{d \text{E}_{\text{Interaction}}}{d \vec{z} }
%
 }
 Assuming the configuration is close to a fixed point, we can write this in terms of the Hessian matrix via a Taylor series expansion and using the fact that the energy, as a function of \textit{z}, is an even function.
  \myequationn{
% \textbf{F}_{\text{Interaction}}
\interactforce
 \paren{z}  = -\frac{d^{2} \interactenergy}{d \vec{z}^{2} } \cdot \vec{z}
 }
 
 At low Reynolds number, any interaction force must be balanced with a viscous force. The viscous force is related to particle velocities by the viscosity $\viscmatrix$, which may include off-diagonal interaction terms.
 \myequationn{
%-\textbf{F}_{\text{Interaction}}
-\interactforce \paren{z}    = \viscforce    =  \viscmatrix \ \cdot \  \dot{\vec{z}}
}

We may now derive the Jacobian:

 \myequationn{
%-\textbf{F}_{\text{Interaction}}
 \dot{\vec{z}} = -\viscmatrix^{-1} \interactforce \paren{z} =  -\viscmatrix^{-1} \ \  \frac{d^{2} \interactenergy}{d \vec{z}^{2} } \cdot \vec{z} = M_{\text{Jacobian}} \vec{z}
}
where:
\myequationn{
M_{\text{Jacobian}} =  -\viscmatrix^{-1} \ \  \frac{d^{2} \interactenergy}{d \vec{z}^{2} } 
}
In this paper, we assume that the system takes a steepest descent, i.e. $\viscmatrix \propto \ \text{I}$; this results in the Jacobian being proportional to the Hessian. In reality, viscous interactions between particles will give rise to off-diagonal terms in $\viscmatrix$ and prevent the system from taking an \emph{exact} steepest
descent. The minus sign arises because a positive velocity gives a negative viscous force. Thus the diagonal terms in the viscosity matrix are negative.

\subsection{Trimer Jacobian}
Because the trimer Jacobian is simpler, we will derive it first. 
The energy of a trimer only comes from one bending angle: 
\myequationn{
\text{E}_{\text{Trimer}} = -\frac{1}{2} {\paren{\theta_{1}}^{2} }
}
%
%\myequationn{
%\text{E}_{\text{Trimer}} = -\frac{1}{2} \paren{\theta_{1}}^{2}
%}
Plugging in:
\myequationn{
\theta_{1} = 2 z_{1} - z_{0} - z_{2}, 
}
we find:
%
%%$\theta_{1} = 2 z_{1} - z_{0} - z_{2}$, we find:
%
\myequationn{
\text{E}_{\text{Trimer}}  = -\frac{1}{2}\cdot  \begin{pmatrix} % or pmatrix or bmatrix or Bmatrix or ...
      + z_{0}^{2} & - 2z_{0} \cdot z_{1} & +  z_{0}\cdot z_{2} \\
    - 2z_{0} \cdot z_{1}   & 4 z_{1}^{2}  & -2 z_{1} \cdot z_{2}\\
      +  z_{0\cdot}z_{2} & -2  z_{1}\cdot z_{2} &  +z_{2}^{2}\\
   \end{pmatrix}.
}
%
%\myequationn{
%\text{E}_{\text{Trimer}}  = -\frac{1}{2}\cdot  \begin{pmatrix} % or pmatrix or bmatrix or Bmatrix or ...
%      + z_{0}^{2} & - 2z_{0} \cdot z_{1} & +  z_{0}\cdot z_{2} \\
%    - 2z_{0} \cdot z_{1}   & 4 z_{1}^{2}  & -2 z_{1} \cdot z_{2}\\
%      +  z_{0\cdot}z_{2} & -2  z_{1}\cdot z_{2} &  +z_{2}^{2}\\
%   \end{pmatrix}.
%}
This allows us to construct the trimer Jacobian ${M_{\text{Trimer}}}$:

\myequationn{
\dot{\vec{z}} \approx -\frac{d \text{E}_{\text{Trimer}}}{d \vec{z}}=  \begin{pmatrix} % or pmatrix or bmatrix or Bmatrix or ...
      1 & - 2  & 1 \\
      -2 & 4  & -2 \\
       1 & -2  & 1\\
   \end{pmatrix} \vec{z}  = {M_{\text{Trimer}}} \cdot \vec{z}
}

${M_{\text{Trimer}}}$ has one non-zero eigenvector of (-1, 2, -1), having an eigenvalue of 6.

\subsection{Quadramer Jacobian}

The energy of a quadramer comes from two bending angles: 
\myequationn{
\text{E}_{\text{Quadramer}} = -\frac{1}{2} \bracket{\paren{\theta_{1}}^{2} + \paren{\theta_{2}}^{2}}.
}

Plugging in:
\myequationn{
\theta_{1} = 2 z_{1} - z_{0} - z_{2}. \qquad \theta_{2} = 2 z_{2} - z_{1} - z_{3},
}
gives:
\myequationn{
\text{E}_{\text{Quadramer}}  =
 -\frac{1}{2}
\underbrace{
\cdot  \begin{pmatrix} % or pmatrix or bmatrix or Bmatrix or ...
      + z_{0}^{2} & - 4z_{0} \cdot z_{1} & + 2 z_{0}z_{2} & + 0 \\
      + 0  & 4 z_{1}^{2}  & -4 z_{1} \cdot z_{2} & + 0\\
      + 0 & + 0  & + z_{2}^{2}& + 0\\
      + 0 & + 0 & + 0 & + 0 \\
   \end{pmatrix}
   }_{\paren{\theta_{1}}^{2}}
\qquad
-\frac{1}{2}
\underbrace{
\cdot  \begin{pmatrix} % or pmatrix or bmatrix or Bmatrix or ...
      + 0 & + 0 & + 0 & + 0 \\
      +0 & + z_{1}^{2} & - 4z_{1} \cdot z_{2} & + 2 z_{1}z_{3} \\
      +0 & + 0  & 4 z_{2}^{2}  & -4 z_{2} \cdot z_{3}\\
      +0 & + 0 & + 0  & + z_{3}^{2}\\
   \end{pmatrix}.
}_{\paren{\theta_{2}}^{2}}
}
This allows us to construct the quadramer Jacobian ${M_{\text{Quadramer}}}$:

\myequationn{
\dot{\vec{z}} \approx -\frac{d \text{E}_{\text{Quadramer}}}{d \vec{z}} =
\bracket{
\underbrace{
  \begin{pmatrix} % or pmatrix or bmatrix or Bmatrix or ...
      1 & - 2  & 1 & 0  \\
      -2 & 4  & -2 & 0 \\
       1 & -2  & 1 & 0\\
       0 & 0  & 0 & 0\\       
   \end{pmatrix} 
   }_{\text{$\theta_{1}$ Component}}
  +
\underbrace{
  \begin{pmatrix} % or pmatrix or bmatrix or Bmatrix or ...
       0 & 0  & 0 & 0\\       
      0&1 & - 2  & 1  \\
      0&-2 & 4  & -2 \\
       0&1 & -2  & 1\\
   \end{pmatrix} 
   }_{\text{$\theta_{2}$ Component}}
   } \cdot \vec{z}
=   \begin{pmatrix} % or pmatrix or bmatrix or Bmatrix or ...
      1 & -2 &1 & 0 \\
      -2 & 5 & -4 & 1 \\
      1 & -4 & 5 & -2 \\
      0 & 1 & -2 & 1 \\
   \end{pmatrix} \cdot \vec{z}
= {M_{\text{Quadramer}}} \cdot \vec{z}.
}

${M_{\text{Quadramer}}}$ has two non-zero eigenvectors:
\begin{enumerate}
\item
(1, -1, -1, 1), which represents U-Bending, has an eigenvalue of 2.  
\item (1, -3, 3, -1), which represents S-Bending, has an eigenvalue of 10.
\end{enumerate}

\phantom{k}

This predicts that U-Bending will be 5 times slower than S-Bending, and 3 times slower than trimer bending.
%\section{Estimating instability growth rate: A dimensional analysis approach}
%\Sam{Not sure what to do with this section-do we want it or not? Could go either way}
%To estimate the growth rate of the instability using dimensional analysis, we consider the ratio between the "force per length" and the "force per velocity " which, when taking the ratio, has units of 1/time.  The force per length is simply the surface tension and the force per velocity, Stokes' law, is viscosity times thickness of the oil film that envelops the inner drops.That is, 
%\myequationn{
 %\frac{1}{time} = \frac{\text{force per length}}{\text{force per velocity}} = \frac{\text{surface tension}}{\text{viscosity} \cdot \text{thickness}}
%}

\section{Bending Energy with Dimensions}
With appropriate dimensions, the bending energy can be expressed as:\\
\begin{equation*}
E\;=\;-\gamma\;l^2\;(\theta_1^2\;+\;\theta_2^2)
\end{equation*}
where $\gamma$ is the surface tension and $l$ is a characteristic length. For this equation to hold, two important conditions must be met:
\begin{enumerate}
\item
$l$  $<<$  $R$ where $R$ is the radius of the drops, and 
\item $z$ $<<$  $R$ where $z$ is the position of the center of the drops.
\end{enumerate}

For the first condition, we can \textit{determine} a value for $l$ based on experimental parameters. 
 Starting with the energy due to surface tension:\\
 \begin{equation*}
E \;= \; \gamma\; l^2
\end{equation*}
and from the definition of Stoke's viscous drag force divided by velocity:
      \begin{equation*}
       \text{Viscous drag force/velocity}\; = \; \mu\; R 
      \end{equation*}
 from which we can determine the speed:
          \begin{equation*}
         \text{speed}\; =\; \frac{force \times R}{\mu R \times R}\;=\;\frac{energy}{\mu R^2}
      \end{equation*}
 and thus the bending rate is:
    \begin{equation*}
         \text{Bending rate (1/bending time)}\; =\; \frac{speed}{R}\;=\;\frac{\gamma l^2}{\mu R^3}
      \end{equation*}
Thus the characteristic length $l$ can be determined in terms of known experimental parameters:
    \begin{equation*}
        l\; =\; \sqrt[]{\frac{\mu\;R^3}{\gamma\;t}}
      \end{equation*}
Using experimental values where 
\begin{equation*}
\mu \;=\; 0.00164\:\frac{N\cdot s}{m^2}\qquad\text{and}\qquad \gamma\;=\;5.2\:\times\;10^{-3}\;\frac{N}{m}
\end{equation*}

 For U bending: t = 90 ms and R $\approx$ 20 $\mu$m, we get \textit{l} to be 0.167 $\mu$m.\\
 For S bending: t = 16 ms and R $\approx$   10 $\mu$m, we get   \textit{l} to be 0.14 $\mu$m.\\
 Both of these \textit{l} values are much less than the radius of the drops. Note that this hand-waving derivation does \textit{not} take into consideration the number of drops or eigenmodes.\\
 
 For the second condition, we acknowledge that $z$ is not much less than $R$. Nonetheless our model describes our experimental results quite well.\\